\documentclass{emulateapj}

\newcommand{\leff}{\mbox{$\lambda_{\rm eff}$}}
\shorttitle{Globular Cluster in M31}
\shortauthors{Ma et al.}

\begin{document}
\slugcomment{AJ, in press}

\title{Age and structure parameters of a remote M31 globular cluster B514 based on {\sl HST}, 2MASS, {\it GALEX} and
BATC observations}
\author{
Jun Ma\altaffilmark{1,2}, Song Wang\altaffilmark{1,3}, Zhenyu Wu\altaffilmark{1},
Zhou Fan\altaffilmark{1}, Tianmen Zhang\altaffilmark{1}, Jianghua Wu\altaffilmark{1},
Xu Zhou\altaffilmark{1}, Zhaoji Jiang\altaffilmark{1} and Jiansheng Chen\altaffilmark{1}}

\altaffiltext{1}{National Astronomical Observatories, Chinese Academy of Sciences, Beijing 100012,
P.R. China; majun@nao.cas.cn}

\altaffiltext{2}{Key Laboratory of Optical Astronomy, National Astronomical Observatories, Chinese
Academy of Sciences, Beijing, 100012, China}

\altaffiltext{3}{Graduate University of Chinese Academy of Sciences, A19 Yuquan Road, Shijingshan
District, Beijing 100049, China}

\begin{abstract}
B514 is a remote M31 globular cluster which locating at a projected distance of $R_p\simeq 55$ kpc. Deep observations with the Advanced Camera for Surveys (ACS) on the {\sl Hubble Space Telescope (HST)} are used to provide the accurate integrated light and star counts of B514. By
coupling analysis of the distribution of the integrated light with star counts, we are able to reliably follow the profile of the cluster out to $\sim 40\arcsec$. Based on the combined profile, we study in detail its surface brightness distribution in F606W and F814W filters,
and determine its structural parameters by fitting a single-mass isotropic King model. The results showed that, the surface brightness distribution departs
from the best-fit King model for $r>10''$. B514 is quite flatted in the inner region, and has a larger half-light radius than majority of normal globular clusters of the same luminosity. It is interesting that, in the $M_V$ versus $\log R_h$ plane, B514 lies nearly on the threshold
for ordinary globular clusters as defined by Mackey \& van den Bergh. In addition, B514 was observed as part of the Beijing-Arizona-Taiwan-Connecticut (BATC) Multicolor Sky Survey, using 13 intermediate-band filters covering a wavelength range of 3000--8500 \AA. Based on aperture photometry, we obtain its SEDs as defined by the 13 BATC filters. We determine the cluster's age and mass by comparing its SEDs (from 2267 to 20000{\AA}, comprising photometric data in the near-ultraviolet of {\it GALEX}, 5 SDSS bands, 13 BATC intermediate-band, and 2MASS
near-infrared $JHK_{\rm s}$ filters) with theoretical stellar population synthesis models, resulting in age of $11.5\pm3.5$ Gyr. This age confirms the previous suggestion that B514 is an old GC in M31. B514 has a mass of $0.96-1.08\times 10^6 \rm M_\odot$, and is a medium-mass globular cluster in M31.
\end{abstract}

\keywords{galaxies: evolution -- galaxies: individual (M31) --
star clusters: B514}

\section{Introduction}

In hierarchical cosmological models, galaxies are built up through the continual accretion and merger of smaller ones. The signature of these system assembly processes is expected to be seen in the outskirts of a galactic halo. Globular clusters (GCs), as luminous compact objects that are found out to distant radii in the haloes of massive galaxies, can serve as excellent tracers of substructures in the outer region of their parent galaxy. So, detailed studies on GCs in the outer haloes of the local galaxies are very important.

M31, with a distance modulus of 24.47 \citep{Holland98,Stanek98,McConnachie05}, is an ideal
nearby galaxy for studying GCs, since it is so near, and contains more GCs than all other Local Group galaxies combined \citep{battis87,raci91,harris91,fusi93}. The study of GCs in M31 was initiated by \citet{hubble32}, who discovered 140 GC candidates with $m_{pg}\leq 18$ mag. And
then, a number of catalogs of GC candidates were published. For example, the Bologna Group
\citep{battis80,battis87,battis93} did independent searches of GC candidates and compiled them with their own Bologna number. Bologna catalog contains a total of 827 objects, and all the objects were classified into five classes by authors' degree of confidence. 353 of these candidates were considered as class A or class B by high level of confidence, and the others fell into class C, D,
or E. {\it V} magnitude and $B-V$ color for most candidates were given in the Bologna catalog. There are recent works dealing with the searched and the catalogs of M31 GCs \citep[e.g.,][]{moche98,bh01,gall04,gall06,gall07,huxor05,huxor08,huxor11,kim07,mackey06,mackey07,
mackey10,Martin06,cald09,cald11,peacock09}.
The continued importance of the study of GCs in this galaxy has been reviewed by \citet{bh00}.

From the spatial structure and internal stellar kinematics of GCs, we can get information on both their formation conditions and dynamical evolution within the tidal fields of their host galaxies. For example, the structural parameters of GCs indicate on what timescales the cluster is ¡°bound¡± to dissolve. However, the integrated properties of GCs, such as age and metallicity, are believed to reflect conditions in the early stages of galaxy formation (Brodie \& Strader 2006).

The most direct method to determine a cluster's age is by employing main-sequence photometry, since the absolute magnitude of the main-sequence turnoff is predominantly affected by age \citep[see][and references therein]{puzia02b}. However, until recently \citep[cf.][]{perina09}, this method was only applied to the star clusters in the Milky Way and its satellites
\citep[e.g.,][]{rich01}, although \citet{brown04} estimated the age of an M31 GC using extremely deep images observed with the {\sl HST}'s Advanced Camera for Surveys (ACS). Generally, the ages of extragalactic star clusters are determined by comparing their observed spectral-energy
distributions (SEDs) and/or spectroscopy with the predictions of SSP models
\citep{Williams01a,Williams01b,degrijs03a,degrijs03b,degrijs03c,bik03,jiang03,Beasley04,puzia05,
fan06,ma06a,ma07a,ma09a,ma09b,ma11,cald09,cald11,Wang10,Perina11}.

M31 GC B514 (B for `Bologna', see Battistini 1987), which was detected by \citet{gall05} based on the XSC sources of the All Sky Data Release of the Two Micron All Sky Survey (2MASS) within a $\sim 9^\circ \times 9^\circ$ area
centered on M31, is the outermost cluster known in M31 at that time, which locating at a projected distance of $R_p\simeq 55$ kpc. Now, many new members of M31 halo GC system, which extending to very large radii, have been discovered \citep[e.g.,][]{huxor05,huxor08,huxor11,mackey06,mackey07,mackey10,Martin06}.

\citet{gall06} presented deep color-magnitude diagram for B514 in F606W and F814W photometry obtained with the ACS/{\sl HST}, which reveals a steep red giant branch (RGB) and a horizontal branch (HB) extending blueward of the instability strip showing that B514 is a classical old
metal-poor GC. \citet{Federici07} studied the density profile of B514 based on the {\sl HST}/ACS observations as \citet{gall06} had, and they found that the light and the star-count profiles show a departure from the best-fitted empirical models of \citet{king62} for $r\ge 8''$ -- as a
surface brightness excess at large radii, and the star-count profile shows a clear break in the
correspondence of the estimated tidal radius; they also found that B514 has a significantly larger
half-light radius than ordinary GCs of the same luminosity. \citet{Clementini09} identified a rich harvest of RR Lyrae stars in B514, based on {\sl HST} Wide Field Planetary Camera 2 (WFPC2) and {\sl HST}/ACS time-series observations.

Since B514 is located in the halo of M31, i.e., far away from the galaxy's disk, it is (for all practical purposes) only affected by the Galactic foreground extinction.
About the foreground Galactic reddening in the direction of M31, it was discussed by many authors \citep[e.g.,][]{vand69,McRa69,fpc80,fusi05}, and nearly similar values were determined such as $E(B-V)=0.08$ by van den Bergh (1969), 0.11 by \citet{McRa69} and \citet{hodge92}, 0.08 by \citet{fpc80}. We argue that the reddening value of B514 should not smaller than the foreground Galactic reddening in the direction of M31. In this paper, we adopt the reddening value of $E(B-V)=0.10$ from \citet{gall06}. The reddening law from \citet{car89} is employed in this paper. In addition, throughout this paper we adopt a distance to M31 of $783 \pm 25$ kpc ($1{\arcsec}$ subtends 3.8 pc), corresponding to a distance modulus of $(m-M)_0 = 24.47 \pm 0.07$ mag \citep{McConnachie05}.

In this paper, we describe the details of the observations and our approach to the data reduction with the {\sl HST}/ACS and the BATC system in \S 2 and \S 3. We will study the surface distribution of B514 with the King models of \citet{king66} in details, which are developed by \citet{michie63} and \citet{king66} based on the assumption that the GCs are formed by single-mass, isotropic, lowered isothermal spheres (hereafter `King models'). We determine the age and mass of B514 by comparing observational SEDs with population synthesis models in \S 4. We provide a summary in \S 5.

\section{Observation and photometric data with {\sl HST}/ACS}

The images of B514 used in this paper were observed with the ACS/Wide Field Camera (WFC) in the F606W and F814W filters on 2005 July 19 (program ID GO 10394, PI: N. Tanvir), covering the period 2005, July 19--20 in F606W (total $t_{\rm exp} = 1776$ s) and F814W (total $t_{\rm exp} = 2505$ s). Upon retrieval from the STScI archive, all images were processed by the standard ACS calibration
pipeline, in which bias and dark subtractions, flatfield division, and the masking of known bad pixels are included. Subsequently, photometric header keywords are populated. In the final stage of the pipeline, the MultiDrizzle software is used to correct the geometric distortion presented in
the images. Finally, any cosmic rays are rejected while individual images in each band are combined into a final single image. We checked the images, and did not find saturated cluster stars. Figure 1 show the images observed with the ACS/WFC in the F606W and F814W.
The ACS/WFC spatial resolution is $0.05\arcsec$ pixel$^{-1}$.

\subsection{Ellipticity, position angle, and surface brightness profile}

Surface photometry of the cluster is obtained from the drizzled images, using the {\sc iraf} task {\sc ellipse}. Its center position was fixed at a value derived by object locator of {\sc ellipse} task, however an initial center position was determined by centroiding. Elliptical isophotes were fitted to the data, with no sigma clipping. We ran two passes of {\sc ellipse} task, the first pass was run in the usual way, with ellipticity and position angle allowed to vary with the isophote semimajor
axis. In the second pass, surface brightness profiles on fixed, zero-ellipticity isophotes were measured, since we choose to fit circular models for the intrinsic cluster structure and the point spread function (PSF) as \citet{barmby07} did (see \S 2.3 for details). The background value was derived as the mean of a region of $100\times100$ pixels in ``empty'' areas far away from the
cluster. We performed the photometric calibration using the results of \citet{sirianni05}: 26.398 in F606W and 25.501 in F814W zero-points. Magnitudes are derived in the ACS/WFC {\sc vegamag} system.

Tables 1 and 2 list the ellipticity, $\epsilon=1-b/a$, and the position angle (P.A.) as a function of the semi-major axis length, $a$, from the center of annulus in the F606W and F814W filters, respectively. These observables have also been plotted in Figure 2; the errors were generated by the {\sc iraf} task {\sc ellipse}, in which the ellipticity errors are obtained from the internal errors in the harmonic fit, after removal of the first and second
fitted harmonics. From Table 1 and Figure 2, we can see that, the values of ellipticity and position angle cannot be obtained beyond $0.9744\arcsec$ in the F606W filter because of low signal-to-noise ratio. In addition, Figure 2 shows that the ellipticity varies significantly with
position along the semimajor axis radius smaller than $\sim 0.1752\arcsec$. Beyond $\sim 0.1752\arcsec$, the ellipticity does not vary significantly as a function of the cluster¡¯s semimajor axis. The P.A. does not vary as a function of the cluster¡¯s semimajor axis within $\sim
0.1752\arcsec$ because of high signal-to-noise ratio; however, beyond this position, it varies significantly with great errors because of low signal-to-noise ratio.

Tables 3 and 4 list the surface brightness profile, $\mu$, of B514, and its integrated magnitude, $m$, as
a function of radius in the F606W and F814W filters, respectively. The errors in the surface brightness were generated by the {\sc iraf} task {\sc ellipse}, in which they are obtained directly from the root mean square scatter of the intensity data along the zero-ellipticity
isophotes. In addition, the surface photometries at radii where the ellipticity and position angle cannot be measured, are obtained based on the last ellipticity and position angle as the {\sc iraf} task {\sc ellipse} is designed.

In order to derive the surface brightness profile of B514 in its outer region, we use the profile from star counts. We used the DOLPHOT\footnote{http://americano.dolphinsim.com/dolphot/}
photometry software (Dolphin 2000a), specifically the ACS module, to photometer our images. DOLPHOT performs PSF fitting using PSFs especially tailored to the ACS camera. Photometry was done simultaneously on all the flat-fielded images from the STScI archive (both filters),
relative to a deep reference image--we used the drizzled combination of the F606W image. DOLPHOT accounts for the hot-pixel and cosmic-ray masking information attached to each flat-fielded image, fits the sky locally around
each detected source, and automatically
applies the correction for the charge-transfer efficiency (CTE, Dolphin 2000b). It then transforms instrumental magnitude to the {\sc vegamag} system (Dolphin 2000b). A variety of quality information is listed with each detected object, including the object type (stellar, extended,
etc.), $\chi^2$ of the PSF fit, sharpness and roundness of the object, and a ``crowding'' parameter which describes how much brighter an object would have been had neighboring objects not been fitted simultaneously. We used the quality information provided by DOLPHOT to clean the resulting detection lists, selecting only stellar detections, with valid photometry on all input images, global sharpness parameter between $-0.3$ and 0.3 in each filter, and crowding parameter less than 0.25 in each filter.

We joined the two profiles into one based on the method of \citet{Federici07}. This involves matching the intensity scales of the two profiles by fitting both profiles to smooth curves in the region $r = 9-16''$. The star count profile is listed in Table 5. The errors for the star counts take account of Poisson statistical uncertainties. The joined profile covered the full $0<r\le40''$ range, as
shown in Figure 3.

\begin{figure*}
\centering
\includegraphics[width=0.65\textwidth]{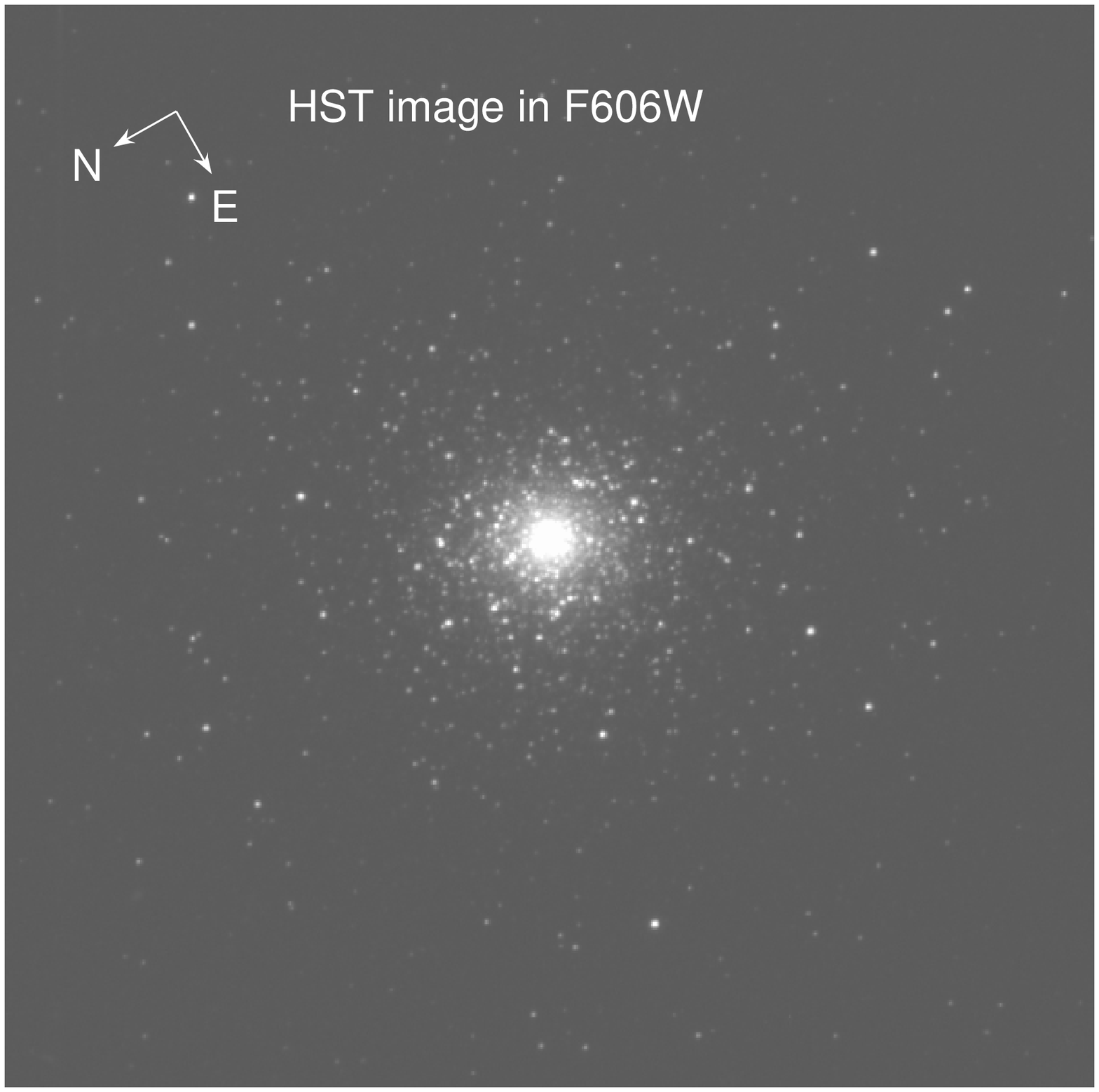}\\
\includegraphics[width=0.65\textwidth]{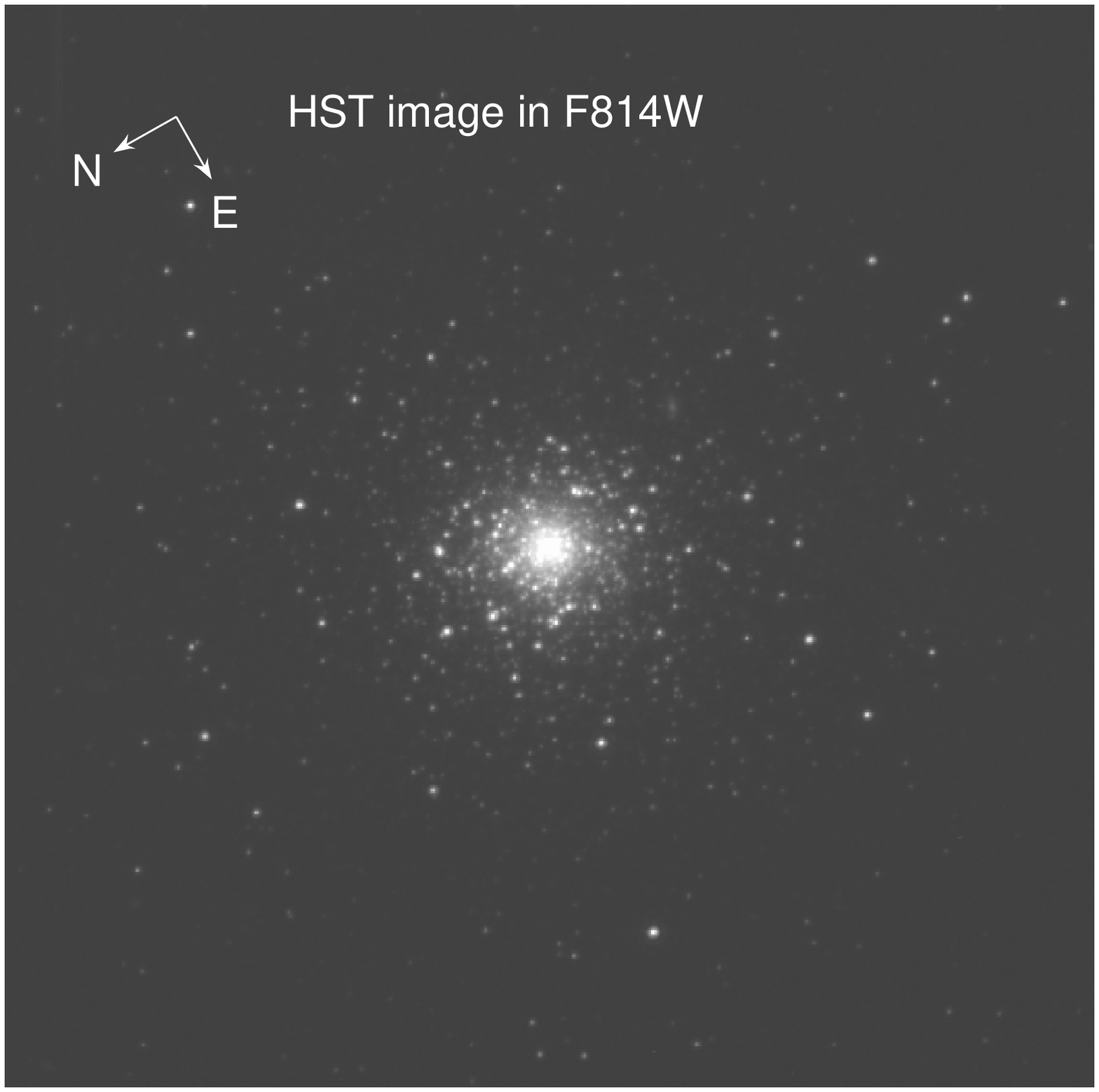}
\vspace{0.5cm}
\caption{The images of GC B514 observed in the F606W and F814W filters of ACS/{\sl HST}. The image
size is $20\arcsec\times20\arcsec$ for each panel.} \label{fig1}
\end{figure*}

\begin{figure*}
\centering
\includegraphics[height=150mm,angle=-90]{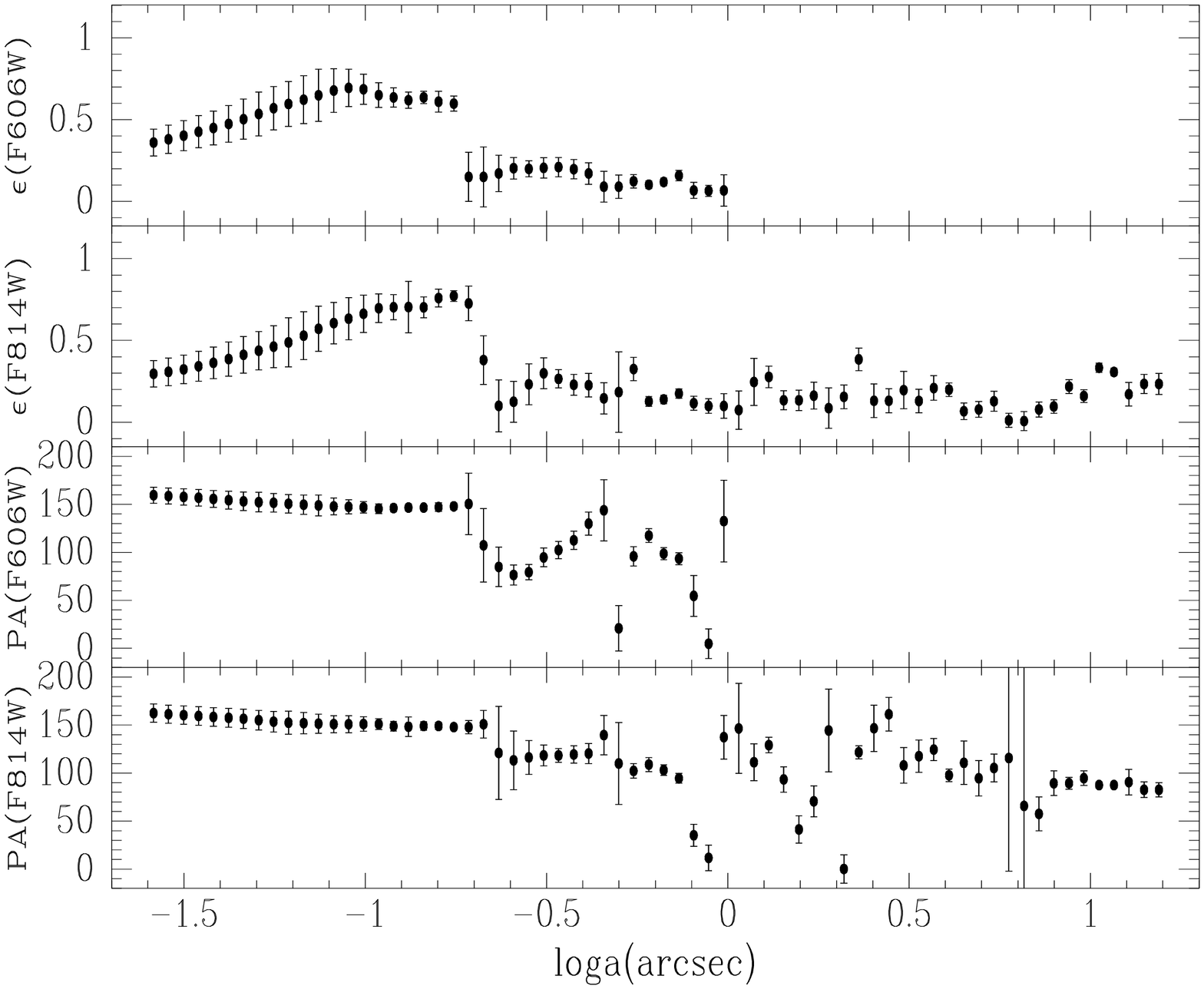}%
\caption{\label{Fig:maps}Ellipticity and P.A. as a function of the semimajor axis in the F606W and F814W
filters of ACS/{\sl HST}.}
\end{figure*}

\subsection{Point spread function}

At a distance of 783 kpc, the ACS/WFC has a scale of $\rm {0.05''=0.19~pc~pixel^{-1}}$, and thus M31 clusters are clearly resolved with it. Their observed core structures, however, are still affected by the PSF. We chose not to deconvolve the data, instead fitting structural models after convolving them with a simple analytic description of the PSF as \citet{barmby07} and \citet{mclaughlin08} did \citep[see][for details]{barmby07,mclaughlin08,ma11}. In addition, since this PSF formula is radially symmetric and the models of \citet{king66} we fit are intrinsically spherical, the convolved models to be fitted to the data are also circularly symmetric.

\subsection{Models and fits}

\subsubsection{Structural models}

After elliptical galaxies, GCs are the best understood and most thoroughly modelled class of stellar systems. For example, a large majority of the $\sim 150$ Galactic GCs have been fitted by the simple models of single-mass, isotropic, lowered isothermal spheres developed by
\citet{michie63} and \citet{king66} (i.e. King models), yielding comprehensive catalogs of cluster structural parameters and physical properties \citep[see][and references therein]{McLaughlin05}. For extragalactic GCs, {\sl HST} imaging data have been used to fit King
models to a large number of GCs in M31 \citep[e.g.,][and references therein]{bk02,barmby07,barmby09}, in
M33 \citep{Larsen02}, and in NGC 5128 \citep[e.g.,][and references therein]{harris02,mclaughlin08}. In addition, there are other models used to
fit the surface profile of GCs, including \citet{wilson75}, \citet{elson87}, and \citet{sersic68}. In this paper, we fit King models to the density profile of B514 observed with ACS/WFC.

\subsubsection{Fits}

Our fitting procedure involves computing in full large numbers of King structural models, spanning a wide range of fixed values of the appropriate shape parameter $W_0$ \citep[see][in detail]{McLaughlin05}. And then the models are convolved with the ACS/WFC PSF for the F606W and
F814W filters \citep[see][for details]{barmby07}:


${\widetilde{I}_{\rm mod}^{*} (R | r_0) = \int\!\!\!\int_{-\infty}^{\infty}
               \widetilde{I}_{\rm mod}(R^\prime/r_0) \times
               \widetilde{I}_{\rm PSF}
                    \left[(x-x^\prime),(y-y^\prime)\right]}$
\begin{equation}
       \ dx^\prime \, dy^\prime\ ,
\label{eq:convol}
\end{equation}
where $\widetilde{I}_{\rm mod}\equiv I_{\rm mod}/I_0$ \citep[see][in detail]{mclaughlin08}. We changed the luminosity density to surface brightness $\widetilde{\mu}_{\rm
mod}^{*}=-2.5\,\log\,[\widetilde{I}_{\rm mod}^{*}]$ before fitting them to the observed surface brightness profile of B514, $\mu=\mu_{0}-2.5\,\log\,[I(R/r_0)/I_0]$, finding the radial scale $r_0$ and central surface brightness $\mu_{0}$ which minimizing $\chi^2$ for every given value
of $W_0$. The $(W_0,r_0,\mu_{0})$ combination that yields the global minimum $\chi_{\rm min}^2$ over the grid used defines the best-fit model of that type:

\begin{equation}
\chi^2  =
  \sum_i{
 \frac{\left[\mu_{\rm obs}(R_i)
             - \widetilde{\mu}_{\rm mod}^{*}(R_i | r_0)
             \right]^2}
      {\sigma_i^2}
        } ,
\label{eq:chi2}
\end{equation}
in which $\sigma_i$ is the error in the surface brightness. Estimates of the one-sigma uncertainties on these basic fit parameters follow from their extreme values over the subgrid of fits with $\chi^2/\nu\le \chi_{\rm min}^2/\nu+1$, here $\nu$ is the number of free parameters. Figure 3 shows our best King fits to B514. In Figure 3, open squares are {\sc ellipse} data points included in the least-squares model fitting, and the asterisks are points not used to constrain the fit; black circles are star-counts points included in the $\chi^2$ model fitting, and red circles are star-counts points not used to constrain the fits. These observed data points shown by asterisks are included in the radius of $R<2~\rm{pixels}=0\farcs1$, and the isophotal intensity is dependent on its neighbors. As \citet{barmby07} pointed out that, the {\sc ellipse} output contains brightnesses for 15 radii inside 2 pixel, but they are all measured from the same 13 central pixels and are not statistically independent. So, to avoid excessive weighting of the central regions of B514 in the fits, we only used intensities at radii $R_{\rm min}$, $R_{\rm min}+(0.5,1.0,2.0)~{\rm pixels}$, or $R>2.5$ as \citet{barmby07} used. Table 6 summarizes the results obtained in this paper.

From Figure 3, we can note that the surface brightness distribution departs from the best-fit King model for $r>10''$, which can be interpreted as the presence of a population of extratidal stars around the cluster. In fact, \citet{Federici07} have reported this population
of extratidal stars (see their Fig. 5 and their discussions).

\begin{figure*}
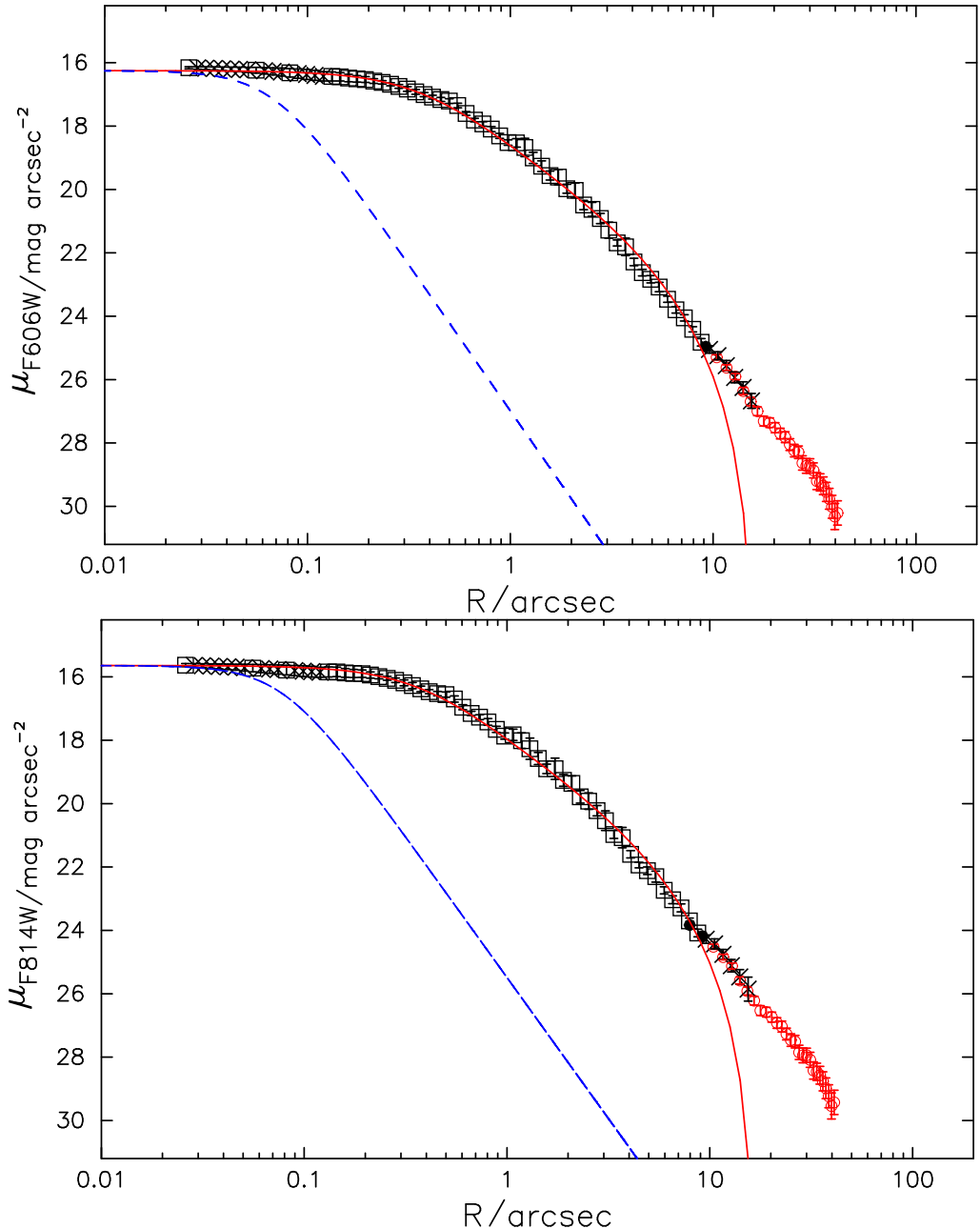

\centering
\includegraphics[width=0.75\textwidth]{f3a.eps}\\
\includegraphics[width=0.75\textwidth]{f3b.eps}
\vspace{0.5cm}
\caption{\label{Fig:spectra}Surface brightness profile of B514 measured in the F606W and F814 filters. Dashed curves (blue) trace the PSF intensity profiles and solid (red) curves are the PSF-convolved best-fit models. Open squares are {\sc ellips} data points and black circles
are star-counts profiles included in the $\chi^2$
model fitting, and the asterisks are {\sc ellips} data points and red circles are star-counts profiles
not used to constrain the fits (see the text in
detail).}
\end{figure*}

\begin{figure*}
\figurenum{4} \epsscale{1.0} \hspace{-1.0cm}\rotatebox{-90}{\plotone{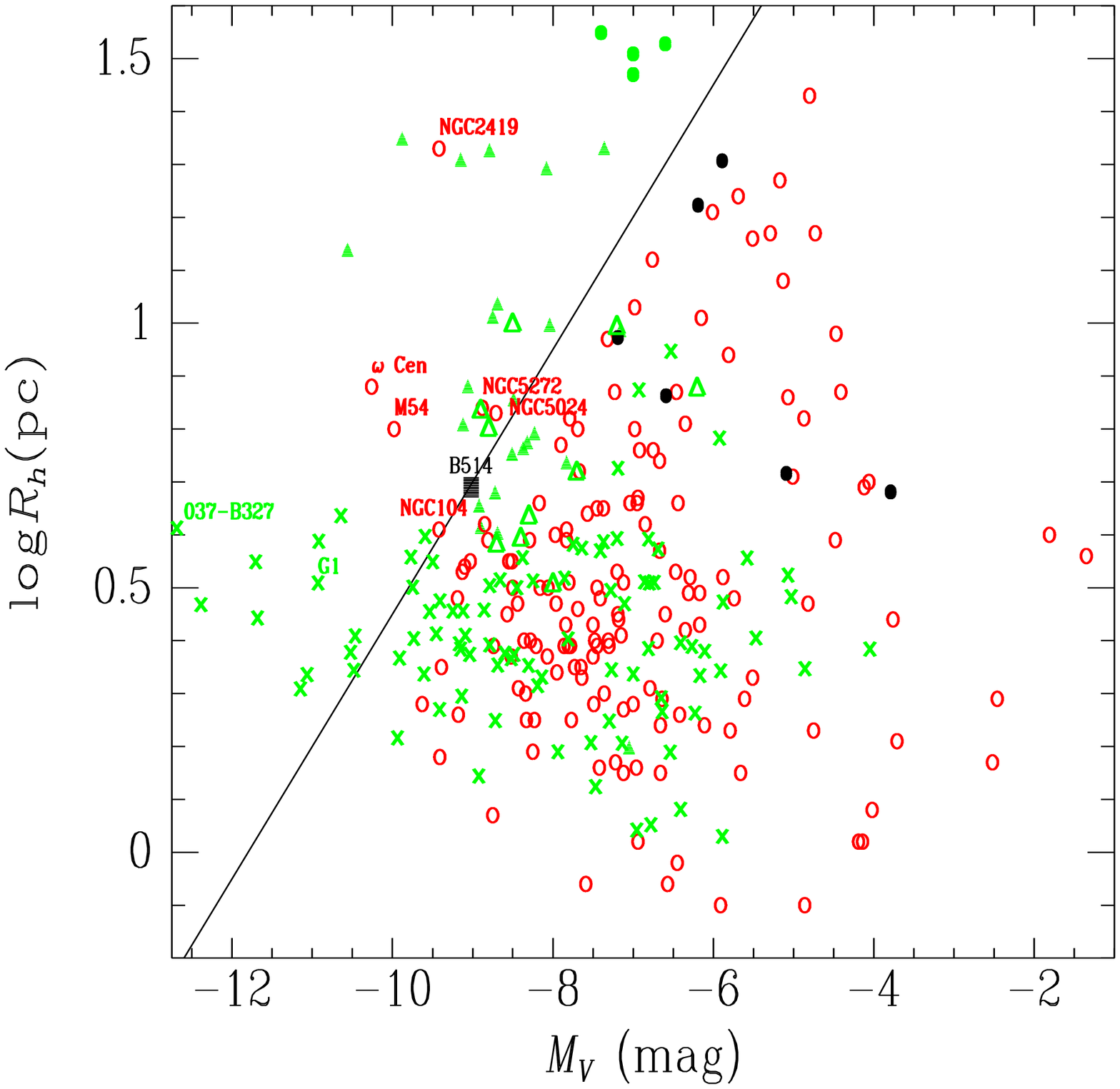}} \vspace{-0.5cm}
\caption{$M_V$ vs. $R_h$ for interesting stellar systems. The plotted line is the threshold for ordinary cluster in this plane as defined by \citet{mackey05}, $\log R_{h} ({\rm pc}) = 0.25 M_{V} ({\rm mag}) + 2.95$. Red circles are Galactic globular clusters from the on-line data base of Harris (1996) (2010 update); green crosses are M31 globular clusters from \citet{barmby07}; green
filled triangles are M31 young massive clusters from \citet{barmby09}; green filled circles are
M31 extended clusters from \citet{mackey06}; green open triangles are outer M31 GCs from \citet{mackey07}; black filled circles are M33 outer halo clusters from \citet{Cockcroft11}; the black square is B514 derived here.}\label{fig4}
\end{figure*}

\subsection{Distribution of B514 in the $M_V$ vs. $\log R_h$ plane}

The distribution of stellar systems in the $M_V$ vs. $\log R_h$ plane can provide interesting information on the evolutionary history of these objects \citep[e.g.,][]{bergh04,mackey05}. In this plane, the half-light radius is an important parameter, which can be used to trace the initial size of a cluster, since it changes little in evolution process \citep[see][for
details]{spitzer72,henon73,lightman78,murphy90}.

Recently, \citet{bergh04} and \citet{mackey05} showed that in a plot of luminosity versus half-light radius, the overwhelming majority of normal Galactic GCs lie below (or to the right) of the line:

\begin{equation}
\log R_{h} ({\rm pc}) = 0.25 M_{V} ({\rm mag}) + 2.95.
\end{equation}

Exceptions to this rule are massive clusters, such as M54 and $\omega$ Centauri in the Milky Way, and G1 in M31, which are widely believed to be the remnant cores of now defunct dwarf galaxies \citep{zinnecker88,freeman93,meylan01}.
Because the well-known giant GC NGC 2419 \citep{bergh04} in the Galaxy and 037-B327 \citep{ma06b} in M31 also lie above this line, it has been speculated that these two objects might also be the remnant cores of dwarf galaxies \citep[but see][for doubts regarding NGC 2419]{degrijs05}.

With the value of $R_h$ (i.e. $r_h$) in the F606W filter obtained in this paper, we plot the relationship of $M_V$ versus $\log R_h$ in Figure 4, in which $M_V=-9.02$ which being derived based on $m_V=15.76$ from \citet{huxor08}. It is interesting that, on this plot B514 is seen to lie nearly on the line defined by equation (3). Considering the uncertainties of $R_h$ and $M_V$, a certain conclusion may not be presented here. However, we argued that, B514 is a medium-mass GC in M31 (see \S 4.4 for details), and is not as massive as G1 and 037--B327 \citep[see][for details]{ma06a,ma06b,ma07b,ma09b}. Furthermore, and
for completeness, in Figure 4 we have also included GCs in the Milky Way, M31 and M33. Galactic GCs are from the on-line data base of Harris (1996) (2010 update). This new revision of the McMaster catalog of Galactic GCs is the first update since 2003 and the biggest single revision since the original version of the catalog published in 1996. The starting points for the present list of structural parameters are the major compilations of
\citet{McLaughlin05} and \citet{Trager95}. \citet{McLaughlin05} used the same raw data as
\citet{Trager95}, and derived structural parameter values from \citet{king66} dynamical profile models. M31 GCs are from recent compilations of data by \citet{barmby07,barmby09}, \citet{mackey06,mackey07}. M33 GCs are from \citet{Cockcroft11}. \citet{barmby07} derived structural parameters for 34 GCs in M31 based on ACS/{\sl HST} observations, and the
derived structural parameters are combined with corrected versions of those measured in an earlier survey in order to construct a comprehensive catalog of structural and dynamical parameters for 93 M31 GCs. \citet{barmby09} measured structural parameters for 23 bright young clusters in M31 based on the {\sl HST}/WFPC2 observations, and
suggested that on average they are larger and more concentrated than typical old clusters. \citet{mackey06} determined structural parameters for 4 extended, luminous globular clusters in the outskirts of M31 based on ACS/{\sl HST} observations. These objects were discovered by
\citet{huxor05} and \citet{Martin06}. \citet{mackey07} derived structural parameters for 10 classical GCs in the far outer regions of M31 based on ACS/WFC observations. \citet{Cockcroft11} searched for outer halo star clusters in M33 as part of the Pan-Andromeda
Archaeological Survey using the images taken with the Canada-France-Hawaii Telescope (CFHT)/MegaCam, and found one new unambiguous star cluster in addition to the five previously known in the M33 outer halo, and determined structural parameters for these 6 outer halo clusters.

From Figure 4, we can see that, for the data of Galactic GCs which updated in 2010, in addition to the clusters already noted by \citet{mackey05}, i.e. M~54 (NGC~6715), ${\omega{\rm~Cen}}$ (NGC~5139), NGC~2419, there are three other bright GCs (NGC~104, NGC~5272, and NGC~5024) lying
above the ``ordinary globular clusters'' threshold. For M31 star clusters, in addition to G1, there are 26 other bright clusters lying above the line defined by equation (3). All of these objects are classified as GCs \citep{barmby07,mackey07} or bright young clusters
\citep{barmby09}. For 4 extended, luminous GCs in the outskirts of M31, they all lie above the line defined by equation (3).

Based on F606W and F814W images of B514 obtained with the ACS/{\sl HST} (program ID GO 10565, PI: S. Galleti), \citet{Federici07} also studied in detail its surface brightness distribution in F606W and F814W filters, and determine its structural parameters by fitting a \citet{king62} model to a surface brightness profile.
Comparing the results of \citet{Federici07} with Table 6 of this paper, we find that our model fits produce smaller tidal radii, which resulting in smaller half-light, or effective, radii of a model. In addition, \citet{Federici07} adopted $M_V=-9.1$ being brighter than $M_V=-9.02$ adopted here. So, in \citet{Federici07},
B514 lied above and brightward of the line defined
by equation (3).

\section{Archival images of the BATC Multicolor Sky Survey, 2MASS and {\it GALEX} and photometric data of SDSS}

In this section, we will determine the magnitudes of B514 based on the archival images of the BATC Multicolor Sky Survey, 2MASS and {\it GALEX} using a standard aperture photometry approach, i.e., the {\sc phot} routine in {\sc daophot} \citep{stet87}. In addition, we will introduce the photometric data of B514 from the Sloan Digital Sky Survey (SDSS) obtained by \citet{peacock09}

\subsection{Intermediate-band photometry of B514}

Observations of B514 were also obtained with the BATC 60/90cm Schmidt telescope located at the Xinglong
station of the National Astronomical Observatory of China (NAOC). This telescope is equipped with 15 intermediate-band filters covering the optical wavelength range from 3000 to 10000 \AA \citep[see][for details]{fan09}. Figure 5 shows a finding chart of B514 in the BATC $b$ band (centered at 5795 \AA).

\begin{figure*}
\figurenum{5} \epsscale{1.0} \hspace{1.2cm}\rotatebox{0}{\plotone{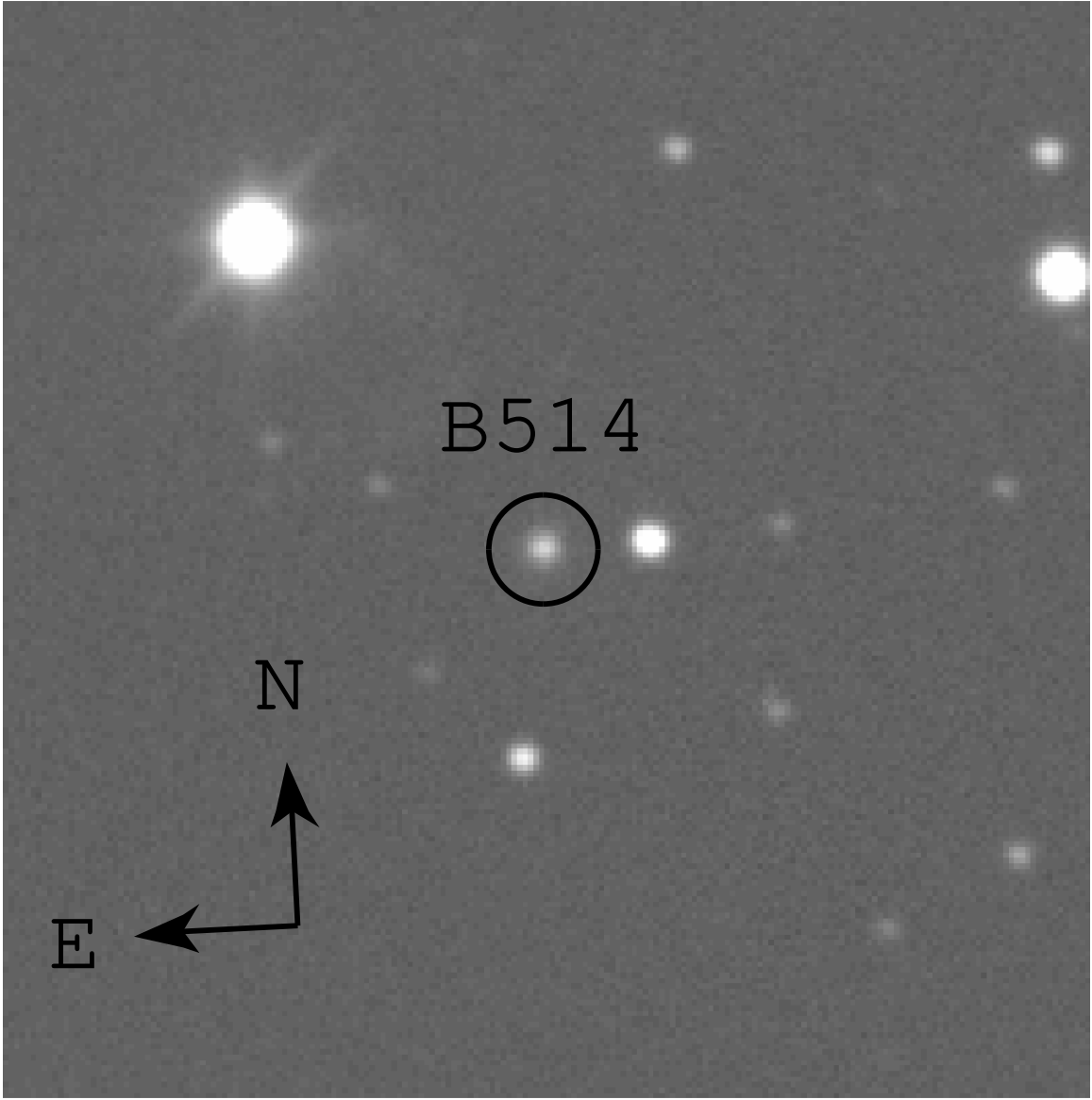}} \vspace{0.5cm}
\caption{Image of B514 in the BATC $b$ band, obtained with the NAOC 60/90cm Schmidt telescope. B514 is circled using an aperture with a radius of $13''$. The field of view of the image is $4.3^{\prime}\times 4.3^{\prime}$.}
\label{fig:6}
\end{figure*}

The BATC survey team obtained 47 images of B514 in 13 BATC filters between 2005 March 1 and 2006 December 9. Table 7 contains the observation log, including the BATC filter names, the central wavelength and bandwidth of each filter, the number of images observed through each filter, and the total observing time per filter. Multiple images through the same filter were combined to improve image quality (i.e., increase the signal-to-noise ratio and remove spurious signal).

Calibration of the magnitude zero level in the BATC photometric system is similar to that of the
spectrophotometric AB magnitude system. For flux calibration, the Oke-Gunn \citep{ok83} primary
flux standard stars HD 19445, HD 84937, BD +26$^{\circ}$2606, and BD +17$^{\circ}$4708 were
observed during photometric nights \citep{yan00}. Column (6) of Table 7 gives the zero-point errors in magnitude for the standard stars through each filter. The formal errors obtained for these stars in the 13 BATC filters used are $\lesssim0.02$ mag, which implies that we can define photometrically the BATC system to an accuracy
of better than 0.02 mag.

We determined the intermediate-band magnitudes of B514 on the combined images. The (radial) photometric asymptotic growth curves, in all BATC bands, flatten out at a radius of $\sim 13''$. Inspection ensured that this
aperture is adequate for photometry, i.e., B514 does not show any obvious signal beyond this radius. Therefore, we use an aperture with $r\approx13''$ for integrated photometry. Since B514 is located in the M31 halo, contamination from background fluctuations can
be neglected. We adopted annuli for background
subtraction spanning between 14 to $20''$. The calibrated photometry of B514 in 13 filters is summarized in column (7) of Table 7, in conjunction with the $1\sigma$ magnitude uncertainties, which include uncertainties from the calibration errors in magnitude from {\sc daophot}.

\subsection{Near-infrared 2MASS photometry of B514}

B514 was detected by \citet{gall05} based on the XSC sources of the All Sky Data Release of 2MASS
within a $\sim 9^\circ \times 9^\circ$ area centered on M31. In order to obtain accurate photometry for B514 in $JHK_s$, we download the images in $JHK_s$ filters including B514. The image in each filter is combined using 6 frames of 1.3 seconds, so the total exposure time of
image in each filter is 7.8 seconds. The mosaic pixel scale of the final atlas image is resampled
to $1''$ \citep[see][for details]{Skrutskie06}. The relevant zero-points for photometry are
20.9210, 20.7089, and 20.0783 in $J$, $H$, and $K_s$ magnitudes, respectively, which are presented
in photometric header keywords. We use an aperture with $r=13''$ for integrated photometry, and annuli for background subtraction spanning between $14''$ to $19''$. The calibrated photometry of B514 in $J$, $H$, and $K_s$ filters is summarized in Table 8, in conjunction with the $1\sigma$ magnitude uncertainties obtained from {\sc daophot}.

\subsection{{\it GALEX} Ultraviolet photometry of B514}

While the principle science goal of the Galaxy Evolution Explorer ({\it GALEX}, \citet{Martin05,Morrissey07}) has been the study of star formation in the local and
intermediate-redshift universe, nearby galaxies such as M31 have also been surveyed, taking advantage of the wide ($1.2^{\circ}$) field of view of {\it GALEX}. The B514 images were obtained as part of the guest program carried out by {\it GALEX} in two UV bands: far-ultraviolet (FUV) ($\leff=1539$\,\AA, FWHM\,$\approx270$\,\AA), and near-ultraviolet (NUV) ($\leff=2316$\,\AA, FWHM\,$\approx615$\,\AA) with resolution
$4.2''$ (FUV) and $5.3''$ (NUV) \citep{Morrissey07}. The exposure times are 1616 seconds in FUV
and 1704 seconds in NUV. The images are sampled with $1.5''$ pixels. The data was downloaded from
the MAST archive. The relevant zero-points for photometry are 20.08 and 18.82 in NUV and FUV magnitudes, respectively \citep{Morrissey07}. We use an aperture with $r=12''$ for integrated photometry, and annuli for background subtraction spanning between $13.5''$ to $19.5''$.
The calibrated photometry of B514 in NUV and FUV filters is summarized in Table 8. From Table 8, we can see that the $1\sigma$ magnitude uncertainties are great, especially the magnitude uncertainty in FUV is very great (2.3 magnitude), i.e. the signal-to-noise ratios of these images are low, especially the signal-to-noise ratio of the image
in FUV is very low. Since the magnitude uncertainty in FUV is very great, we will not use it when fitting to derive the age of B514 in \S 4.

\subsection{Photometric data of B514 from SDSS}

\citet{peacock09} presented an updated catalog of M31 GCs based on images from the Wide Field Camera (WFCAM) on the United Kingdom Infrared Telescope and from the SDSS, in which $ugriz$ and $K$-band photometry are determined. In this catalog, B514 is named H6 from \citet{huxor08}, and $ugriz$ photometry is presented.

\section{Stellar population of B514}

\subsection{Metallicity of B514}

Cluster SEDs are determined by the combination of their ages and metallicities, which is often referred to
as the age-metallicity degeneracy. Therefore, the age of a cluster can only be constrained accurately if the metallicity is known with confidence, from independent determinations. There exist four metallicity determinations for B514: namely, $\rm{[Fe/H]}=-1.8\pm0.3$ (spectroscopic
from Galleti et al. 2005), $-1.8\pm0.15$ (from the CMD; Galleti et al. 2006), $-2.14\pm0.15$ (from the CMD; Mackey et al. 2007), and $-2.06\pm0.16$ (spectroscopic from Galleti et al. 2009), which are consistent. In order to adopt a reasonable value of metallicity for B514, the mean value of these four independent determinations, i.e. $\rm{[Fe/H]=-1.95}$, is adopted in this paper.

\subsection{Stellar populations and synthetic photometry}

To determine the age and mass of B514, we compared its SEDs with theoretical stellar population synthesis models. The SEDs consist of photometric data in NUV of GALEX,
13 BATC intermediate-band and 2MASS near-infrared $JHK_s$ filters obtained in this paper, and of the photometric data in 5 SDSS filters obtained by \citet{peacock09}.
We will not include the photometric datum in the FUV band when constraining the age of B514 because of its large photometric error (2.3 magnitude), i.e. the photometric datum is not accurate. B514 is a very metal poor GC (see discussions above). So, we use the  SSP models of
\citet{bc03} (hereafter BC03), which have been upgraded from the earlier \citet{bc93,bc96} versions, and now provide the evolution of the spectra and photometric properties for a wide range of stellar metallicities. For example, BC03 SSP models based on the Padova-1994 evolutionary tracks include six initial metallicities, $Z= 0.0001, 0.0004, 0.004, 0.008, 0.02\, (Z_\odot)$, and
0.05, corresponding to ${\rm [Fe/H]}=-2.25$, $-1.65$, $-0.64$, $-0.33$, $+0.09$, and $+0.56$. BC03 provides 26 SSP models (both of high and low spectral resolution) using the Padova-1994 evolutionary tracks, half of which were computed based on the \cite{salp55} IMF with lower and
upper-mass cut-offs of $m_{\rm L}=0.1~M_{\odot}$ and $m_{\rm U}=100~M_{\odot}$, respectively. The
other thirteen were computed using the \cite{chabrier03} IMF with the same mass cut-offs. In addition, BC03 provide 26 SSP models using the Padova-2000 evolutionary tracks which including six partially different initial metallicities, $Z = 0.0004$, 0.001, 0.004, 0.008, 0.019 $(Z_\odot)$, and 0.03, i.e., ${\rm [Fe/H]}=-1.65, -1.25, -0.64, -0.33, +0.07$, and $+0.29$. In this paper, we
adopt the high-resolution SSP models using the Padova-1994 evolutionary tracks to determine the most appropriate age for B514 since its metallicity is $\rm
[Fe/H]=-1.95$, and a Salpeter (1955) IMF is used.
These SSP models contain 221 spectra describing the spectral evolution of SSPs from $1.0\times10^5$ yr to 20 Gyr. The evolving spectra include the contribution of the
stellar component at wavelengths from 91\AA~to $160~\mu$m.

Since our observational data are integrated luminosities through a given set of filters, we convolved the theoretical SSP SEDs of BC03 with the {\it GALEX} NUV, SDSS $ugriz$, BATC $a-n$ and 2MASS $JHK_{\rm s}$ filter response curves to obtain synthetic optical and NIR photometry for comparison \citep[see][for details]{ma09a,ma09b,Wang10,ma11}.

\subsection{Fit results}

We use a $\chi^2$ minimization approach to examine which SSP models are most compatible with the observed SEDs, following

\begin{equation}
\chi^2=\sum_{i=1}^{22}{\frac{[m_{\lambda_i}^{\rm intr}-m_{\lambda_i}^{\rm
mod}(t)]^2}{\sigma_{i}^{2}}},
\end{equation}
where $m_{\lambda_i}^{\rm mod}(t)$ is the integrated magnitude in the $i{\rm th}$ filter of a theoretical SSP at age $t$, $m_{\lambda_i}^{\rm intr}$ represents the intrinsic integrated magnitude in the same filter, and $\sigma_i$ is the magnitude uncertainty, defined as
\begin{equation}
\sigma_i^{2}=\sigma_{{\rm obs},i}^{2}+\sigma_{{\rm mod},i}^{2}+\sigma_{{\rm md},i}^{2}.
\end{equation}
Here, $\sigma_{{\rm obs},i}$ is the observational uncertainty from Tables 7 and 8 of this paper,
and Table 1 of \citet{peacock09}, $\sigma_{{\rm mod},i}$ is the uncertainty associated with the model itself,
and $\sigma_{{\rm md},i}$ is associated with the
uncertainty with the distance modulus adopted here.
\citet{charlot96} estimated the uncertainty
associated with the term $\sigma_{{\rm mod},i}$ by comparing the colors obtained from different
stellar evolutionary tracks and spectral libraries. Following \citet{ma09a}, \citet{ma09b}, \citet{Wang10} and \citet{ma11}, we adopt $\sigma_{{\rm mod},i}=0.05$ mag. For $\sigma_{{\rm md},i}$, we adopt 0.07 from \citet{McConnachie05}.

Before fitting, we obtained the the theoretical SEDs for the metallicity $\rm [Fe/H]=-1.95$ model by interpolation of between ${\rm [Fe/H]}=-2.25$ and $-1.65$ models.

Since the observed magnitudes in the 2MASS photometric systems are given in the Vega system, we transformed them to the AB system for our fits. The photometric offsets in the 2MASS filters between the Vega and AB systems were obtained based on equations (7) and (8) in the manual provided by \citet{bc03} (bc03.ps). The best-reduced
$\chi^2_{\rm min}/\nu=0.8$ is achieved with an age of $11.5\pm3.5$ Gyr (1 $\sigma$ uncertainties), $\nu=21$ is the number of free parameters, i.e., the number of observational data points minus the number of parameters used in the theoretical model. In Figure 6, we show the intrinsic SEDs of B514, the integrated SEDs of the
best-fitting model, and the spectra of the best-fitting model. From Figure 6, we can see that the BC03 SSP models cannot fit the photometric data point in $H$ band as well as the other 21 data points, i.e. the observed magnitude is brighter than the model one in the $H$ band. However, the photometric data point in the $H$ band from \citet{gall05} can be fitted by BC03 SSP models as well as the other 21 data points, and the fitting result (the age of B514) is in agreement with one obtained above ($11.5\pm3.5$ Gyr) within the uncertainty.

much better fitted by BC03 SSP models than the data value derived in this paper, and the fitting results are consistent within the uncertainty.

\subsection{Mass of B514}

We next determined the mass of B514. The BC03 SSP models are normalized to a total mass of $1 M_\odot$ in stars at age $t=0$. The absolute magnitudes (in the Vega system) in $V$, SDSS $ugriz$ and 2MASS $JHK_{\rm s}$ filters are included in the BC03 SSP models. The difference between the intrinsic absolute magnitudes and those given by the model provides a direct measurement of the cluster mass.
To reduce mass uncertainties resulting from photometric uncertainties based on only magnitudes in one filter (in general the $V$ band is used), we estimated the mass of B514 using magnitudes in the $V$, $ugriz$ and $JHK_{\rm s}$ bands. The resulting mass determinations for B514 are listed in Table 9 with their $1\sigma$ uncertainties. From Table 9, we can see that the mass of B514 obtained based on the magnitudes in different filters is consistent except for one in the $H$ band. In fact, the observed magnitude is brighter than the model one in $H$ band (see discussion in \S 4.3). So, the mass of B514 derived based on the magnitude in the $H$ band is more massive than its true one. The mass of B514 derived based on the magnitude in the $H$ band from \citet{gall05} is in agreement with ones derived based on the magnitudes in the other 8 bands. Table 9, we know that the obtained mass of B514 is between $0.96-1.08\times 10^6 \rm M_\odot$ not including one in the $H$ band. Comparing with 037-B327 [$\mathcal{M}_{\rm 037-B327}
\sim 8.5 \times 10^6\rm M_\odot$ \citep{bk02} or $\mathcal{M}_{\rm 037-B327} \sim 3.0 \pm 0.5 \times 10^7\rm M_\odot$ \citep{ma06a}] and G1 [$\mathcal{M}_{\rm G1} \sim (7-17)\times 10^6\rm  M_\odot$ \citep{meylan01} or $\mathcal{M}_{\rm G1} \sim (5.8-10.6)\times10^6\rm M_\odot$ \citep{ma09b}] in M31 and $\omega$ Cen
[$\mathcal{M}_{\omega{\rm~Cen}} \sim (2.9 - 5.1) \times
10^6$M$_\odot$ \citep{meylan02}] in the Milky Way, the most massive clusters in the Local Group, B514 is only a medium-mass globular cluster.

\begin{figure*}
\centering
\figurenum{6} \epsscale{1.0}\hspace{0.0cm}\rotatebox{0}{\plotone{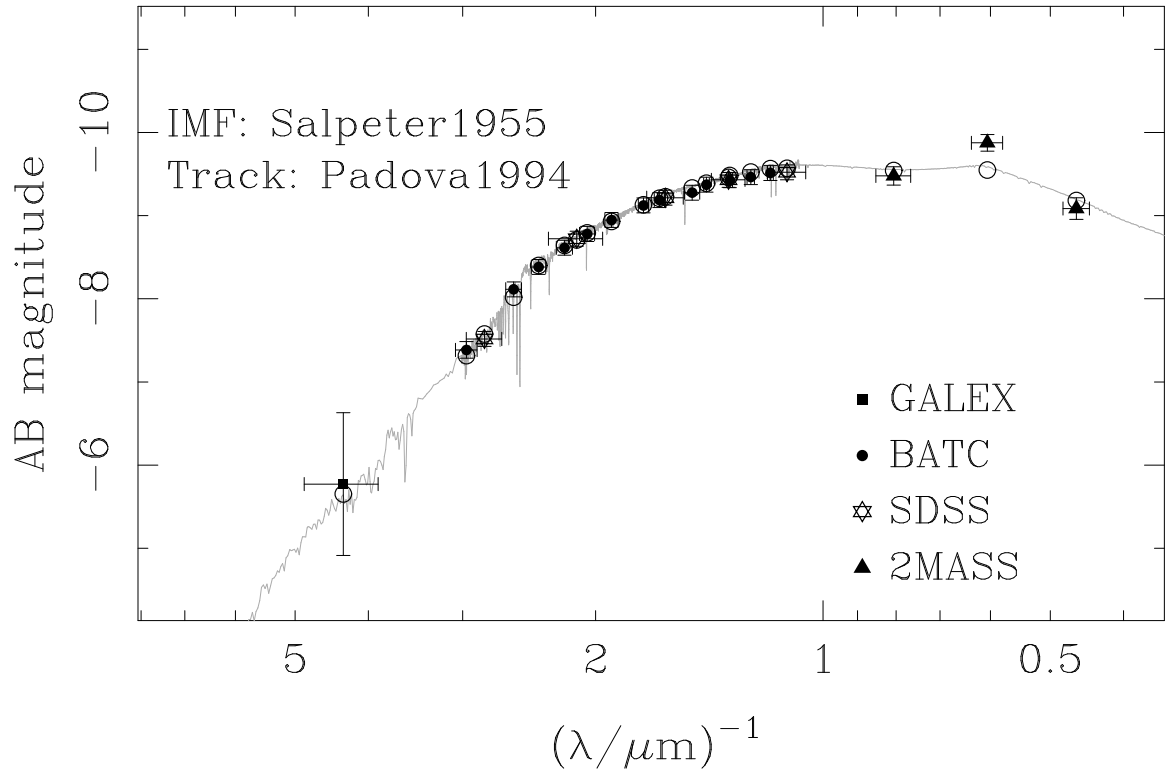}} \vspace{0.5cm}
\caption{Best-fitting, integrated theoretical BC03 SEDs compared to the intrinsic SED of B514. The photometric measurements are shown as symbols with error bars. Open circles represent the calculated magnitudes of the model SED for each filter.} \label{fig:seven}
\end{figure*}

\section{Summary}

In this paper, we determined the structural parameters of one remote globular cluster B514 known in M31 based on F606W and F814W images obtained with the ACS/{\sl HST}. By performing a fit to the surface brightness distribution of a single-mass isotropic King model, we derive its parameters: the best-fitting scale radii $r_0=0.36^{+0.09}_{-0.05}~\rm{arcsec}~(=1.35^{+0.35}_{-0.19}~\rm{pc})$ and $0.36^{+0.09}_{-0.06}~\rm{arcsec}~(=1.35^{+0.32}_{-0.22}~\rm{pc})$, tidal radii
$r_t=16.08^{+2.11}_{-1.35}~\rm{arcsec}~(=61.11^{+8.00}_{-5.14}~\rm{pc})$ and
$16.79^{+1.74}_{-1.48}~\rm{arcsec}~(=63.78^{+6.62}_{-5.64}~\rm{pc})$, and concentration indexes $c=\log
(r_t/r_0)=1.66^{+0.05}_{-0.04}$ and $1.68^{+0.04}_{-0.04}$ in F606W and F814W, respectively; the central surface brightnesses are $16.25^{+0.57}_{-0.56}$ mag arcsec$^{-2}$ and $15.64^{+0.80}_{-0.64}$ mag arcsec$^{-2}$ in F606W
and F814W, respectively; the half-light, or effective, radius of a model that contains half the total luminosity in projection, at $r_h=1.31^{+0.14}_{-0.08}~\rm{arcsec}~(=5.00^{+0.55}_{-0.32}~\rm{pc})$ and $1.36^{+0.13}_{-0.09}~\rm{arcsec}~(=5.17^{+0.48}_{-0.34}~\rm{pc})$
in F606W and F814W, respectively. The results show
that, the surface brightness distribution departs from the best-fit King model for $r>10''$. In addition, B514 was observed as part of the BATC Multicolor
Sky Survey, using 13 intermediate-band filters covering a wavelength range of 3000--80,000 \AA. Based on aperture photometry, we obtain its SEDs as defined by the 13 BATC filters. We determine the cluster's age by comparing its SEDs (from 2267 to 20,000{\AA}, comprising photometric
data in the NUV of {\it GALEX}, 13 BATC intermediate-band filters, and 5 SDSS filters, and 2MASS near-infrared $JHK_{\rm s}$ data) with theoretical stellar population synthesis models, resulting in an age of $11.5\pm3.5$ Gyr. This age confirms previous suggestions that B514 is an old GC in M31. B514 has a mass of $0.96-1.08\times 10^6 \rm M_\odot$, and is a medium-mass globular cluster in M31.

\acknowledgments We would like to thank the anonymous referee for providing rapid and thoughtful report that helped improve the original manuscript greatly. This work is partly based on observations made with the NASA/ESA {\sl Hubble Space Telescope}, obtained at the Space Telescope Science Institute, which is operated by AURA,
Inc., under NASA contract NAS 5-26555. These observations are associated with proposal 10394. This work was supported by the Chinese National Natural Science Foundation grands No. 10873016, 10633020, 10803007, 11003021, and 11073032, and by National Basic Research Program of China (973 Program), No. 2007CB815403.

\clearpage
\begin{table}
\begin{center}
\caption{B514: Ellipticity, $\epsilon$, and position angle (P.A.) as a function of the semimajor axis, $a$, in the F606W filter of {\sl HST} ACS-WFC}
\begin{tabular}{ccc|ccc}
\tableline\tableline
   $a$    &  $\epsilon$  &   P.A.   &   $a$    &  $\epsilon$  &   P.A.    \\
 (arcsec) &              &  (deg)   & (arcsec) &              &   (deg)   \\
\tableline
 0.0260  & $0.360   \pm0.082   $ & $159.5   \pm8.3     $ & 0.1752  & $0.598   \pm0.046   $ & $147.9   \pm3.4     $ \\
 0.0287  & $0.380   \pm0.087   $ & $158.6   \pm8.3     $ & 0.1928  & $0.150   \pm0.150   $ & $150.4   \pm32.0    $ \\
 0.0315  & $0.402   \pm0.092   $ & $157.7   \pm8.5     $ & 0.2120  & $0.150   \pm0.183   $ & $107.3   \pm38.4    $ \\
 0.0347  & $0.426   \pm0.098   $ & $156.9   \pm8.6     $ & 0.2333  & $0.171   \pm0.111   $ & $84.8    \pm20.6    $ \\
 0.0381  & $0.449   \pm0.104   $ & $155.6   \pm8.8     $ & 0.2566  & $0.203   \pm0.066   $ & $76.4    \pm10.5    $ \\
 0.0420  & $0.474   \pm0.112   $ & $154.3   \pm9.2     $ & 0.2822  & $0.199   \pm0.049   $ & $79.3    \pm8.0     $ \\
 0.0461  & $0.503   \pm0.123   $ & $153.1   \pm9.7     $ & 0.3105  & $0.205   \pm0.062   $ & $94.7    \pm9.8     $ \\
 0.0508  & $0.535   \pm0.134   $ & $152.3   \pm10.2    $ & 0.3415  & $0.210   \pm0.059   $ & $102.4   \pm9.1     $ \\
 0.0558  & $0.570   \pm0.133   $ & $151.6   \pm9.7     $ & 0.3757  & $0.197   \pm0.059   $ & $112.5   \pm9.5     $ \\
 0.0614  & $0.596   \pm0.137   $ & $150.6   \pm9.7     $ & 0.4132  & $0.171   \pm0.066   $ & $129.9   \pm12.1    $ \\
 0.0676  & $0.622   \pm0.146   $ & $149.6   \pm10.1    $ & 0.4545  & $0.090   \pm0.094   $ & $143.8   \pm31.9    $ \\
 0.0743  & $0.649   \pm0.159   $ & $148.8   \pm10.8    $ & 0.5000  & $0.090   \pm0.071   $ & $20.8    \pm23.6    $ \\
 0.0818  & $0.678   \pm0.134   $ & $147.8   \pm8.8     $ & 0.5500  & $0.123   \pm0.041   $ & $95.8    \pm10.1    $ \\
 0.0899  & $0.694   \pm0.114   $ & $147.4   \pm7.4     $ & 0.6050  & $0.102   \pm0.024   $ & $117.4   \pm7.2     $ \\
 0.0989  & $0.686   \pm0.092   $ & $146.9   \pm6.0     $ & 0.6655  & $0.119   \pm0.024   $ & $98.6    \pm6.2     $ \\
 0.1088  & $0.650   \pm0.075   $ & $145.4   \pm5.0     $ & 0.7321  & $0.158   \pm0.032   $ & $93.4    \pm6.2     $ \\
 0.1197  & $0.636   \pm0.059   $ & $146.1   \pm4.0     $ & 0.8053  & $0.067   \pm0.049   $ & $54.6    \pm21.3    $ \\
 0.1317  & $0.619   \pm0.050   $ & $146.6   \pm3.6     $ & 0.8858  & $0.065   \pm0.034   $ & $4.8     \pm15.5    $ \\
 0.1448  & $0.636   \pm0.037   $ & $146.6   \pm2.7     $ & 0.9744  & $0.067   \pm0.096   $ & $132.5   \pm42.6    $ \\
 0.1593  & $0.610   \pm0.064   $ & $147.1   \pm4.5     $ &   &  &     \\
\tableline
\end{tabular}
\end{center}
\end{table}
\begin{table}
\begin{center}
\caption{B514: Ellipticity, $\epsilon$, and position angle (P.A.) as a function of the semimajor axis, $a$, in the F814W filter of {\sl HST} ACS-WFC}
\begin{tabular}{ccc|ccc}
\tableline\tableline
   $a$    &  $\epsilon$  &   P.A.   &   $a$    &  $\epsilon$  &   P.A.     \\
 (arcsec) &              &  (deg)   & (arcsec) &              &   (deg)    \\
\tableline
 0.0260  & $0.296   \pm0.081   $ & $162.5   \pm9.5     $ & 0.6655  & $0.139   \pm0.026   $ & $103.1   \pm5.6     $ \\
 0.0287  & $0.308   \pm0.084   $ & $161.4   \pm9.5     $ & 0.7321  & $0.175   \pm0.028   $ & $94.5    \pm5.1     $ \\
 0.0315  & $0.323   \pm0.087   $ & $160.4   \pm9.5     $ & 0.8053  & $0.116   \pm0.043   $ & $35.2    \pm11.4    $ \\
 0.0347  & $0.342   \pm0.091   $ & $159.5   \pm9.6     $ & 0.8858  & $0.099   \pm0.044   $ & $11.7    \pm13.4    $ \\
 0.0381  & $0.363   \pm0.097   $ & $158.5   \pm9.7     $ & 0.9744  & $0.099   \pm0.075   $ & $137.4   \pm22.8    $ \\
 0.0420  & $0.386   \pm0.104   $ & $157.6   \pm9.9     $ & 1.0718  & $0.074   \pm0.117   $ & $146.5   \pm46.9    $ \\
 0.0461  & $0.412   \pm0.112   $ & $156.6   \pm10.1    $ & 1.1790  & $0.246   \pm0.144   $ & $111.3   \pm19.3    $ \\
 0.0508  & $0.437   \pm0.117   $ & $155.1   \pm10.2    $ & 1.2969  & $0.277   \pm0.066   $ & $129.1   \pm8.1     $ \\
 0.0558  & $0.462   \pm0.128   $ & $153.6   \pm10.7    $ & 1.4266  & $0.134   \pm0.058   $ & $93.4    \pm13.2    $ \\
 0.0614  & $0.488   \pm0.150   $ & $152.5   \pm12.0    $ & 1.5692  & $0.134   \pm0.062   $ & $41.3    \pm14.2    $ \\
 0.0676  & $0.529   \pm0.146   $ & $152.0   \pm11.1    $ & 1.7261  & $0.162   \pm0.082   $ & $70.6    \pm16.1    $ \\
 0.0743  & $0.571   \pm0.138   $ & $151.5   \pm10.0    $ & 1.8987  & $0.086   \pm0.124   $ & $144.4   \pm43.2    $ \\
 0.0818  & $0.606   \pm0.127   $ & $151.0   \pm8.9     $ & 2.0886  & $0.155   \pm0.073   $ & $0.1     \pm14.8    $ \\
 0.0899  & $0.633   \pm0.129   $ & $150.9   \pm8.8     $ & 2.2975  & $0.384   \pm0.069   $ & $121.7   \pm6.8     $ \\
 0.0989  & $0.663   \pm0.114   $ & $151.1   \pm7.6     $ & 2.5272  & $0.131   \pm0.103   $ & $146.7   \pm24.2    $ \\
 0.1088  & $0.697   \pm0.087   $ & $151.0   \pm5.6     $ & 2.7800  & $0.131   \pm0.074   $ & $161.3   \pm17.5    $ \\
 0.1197  & $0.703   \pm0.077   $ & $148.9   \pm5.0     $ & 3.0580  & $0.196   \pm0.114   $ & $108.0   \pm18.6    $ \\
 0.1317  & $0.704   \pm0.157   $ & $148.3   \pm10.1    $ & 3.3638  & $0.130   \pm0.072   $ & $117.6   \pm16.9    $ \\
 0.1448  & $0.702   \pm0.064   $ & $149.1   \pm4.4     $ & 3.7001  & $0.209   \pm0.075   $ & $124.5   \pm11.4    $ \\
 0.1593  & $0.759   \pm0.055   $ & $149.1   \pm3.6     $ & 4.0701  & $0.199   \pm0.041   $ & $97.7    \pm6.5     $ \\
 0.1752  & $0.772   \pm0.033   $ & $147.9   \pm2.2     $ & 4.4772  & $0.067   \pm0.051   $ & $110.7   \pm22.6    $ \\
 0.1928  & $0.726   \pm0.106   $ & $147.9   \pm6.9     $ & 4.9249  & $0.078   \pm0.048   $ & $94.6    \pm18.3    $ \\
 0.2120  & $0.380   \pm0.149   $ & $150.9   \pm14.4    $ & 5.4174  & $0.128   \pm0.061   $ & $105.3   \pm14.5    $ \\
 0.2333  & $0.100   \pm0.159   $ & $121.0   \pm48.4    $ & 5.9591  & $0.011   \pm0.043   $ & $115.7   \pm118.0   $ \\
 0.2566  & $0.125   \pm0.125   $ & $113.3   \pm30.5    $ & 6.5550  & $0.007   \pm0.059   $ & $65.8    \pm244.4   $ \\
 0.2822  & $0.232   \pm0.125   $ & $116.3   \pm17.7    $ & 7.2105  & $0.078   \pm0.046   $ & $57.5    \pm17.6    $ \\
 0.3105  & $0.299   \pm0.094   $ & $118.4   \pm10.9    $ & 7.9316  & $0.096   \pm0.041   $ & $89.4    \pm12.8    $ \\
 0.3415  & $0.265   \pm0.056   $ & $118.4   \pm7.2     $ & 8.7247  & $0.218   \pm0.042   $ & $89.4    \pm6.1     $ \\
 0.3757  & $0.229   \pm0.064   $ & $119.4   \pm9.1     $ & 9.5972  & $0.160   \pm0.039   $ & $94.7    \pm7.6     $ \\
 0.4132  & $0.226   \pm0.072   $ & $120.4   \pm10.5    $ & 10.5569 & $0.333   \pm0.029   $ & $87.5    \pm3.0     $ \\
 0.4545  & $0.146   \pm0.096   $ & $139.6   \pm20.5    $ & 11.6126 & $0.307   \pm0.023   $ & $87.5    \pm2.6     $ \\
 0.5000  & $0.184   \pm0.246   $ & $110.0   \pm42.6    $ & 12.7738 & $0.171   \pm0.072   $ & $90.6    \pm13.3    $ \\
 0.5500  & $0.325   \pm0.071   $ & $102.3   \pm7.5     $ & 14.0512 & $0.234   \pm0.058   $ & $82.6    \pm8.1     $ \\
 0.6050  & $0.127   \pm0.030   $ & $108.9   \pm7.3     $ & 15.4564 & $0.234   \pm0.064   $ & $82.6    \pm7.4     $ \\
\tableline
\end{tabular}
\end{center}
\end{table}
\begin{table}
\begin{center}
\caption{B514: Surface brightness, $\mu$, and integrated magnitude, $m$, as a function of the radius in the F606W filter of {\sl HST} ACS-WFC}
\begin{tabular}{ccc|ccc}
\tableline\tableline
   $R$    &  $\mu$                  &   $m$   &   $R$    &  $\mu$                  &   $m$  \\
 (arcsec) &  ($\rm{mag/arcsec^2}$)  &  (mag)  & (arcsec) &  ($\rm{mag/arcsec^2}$)  &  (mag) \\
\tableline
 0.0260  & $16.438  \pm0.034   $ & 22.868   & 0.6655  & $18.091  \pm0.055   $ & 17.008  \\
 0.0287  & $16.445  \pm0.038   $ & 22.868   & 0.7321  & $18.212  \pm0.071   $ & 16.900  \\
 0.0315  & $16.452  \pm0.043   $ & 22.868   & 0.8053  & $18.374  \pm0.080   $ & 16.804  \\
 0.0347  & $16.461  \pm0.048   $ & 22.868   & 0.8858  & $18.588  \pm0.079   $ & 16.705  \\
 0.0381  & $16.470  \pm0.054   $ & 22.868   & 0.9744  & $18.799  \pm0.086   $ & 16.615  \\
 0.0420  & $16.480  \pm0.062   $ & 22.868   & 1.0718  & $18.808  \pm0.134   $ & 16.531  \\
 0.0461  & $16.491  \pm0.070   $ & 22.868   & 1.1790  & $18.953  \pm0.289   $ & 16.434  \\
 0.0508  & $16.503  \pm0.080   $ & 21.221   & 1.2969  & $19.283  \pm0.175   $ & 16.362  \\
 0.0558  & $16.517  \pm0.090   $ & 21.221   & 1.4266  & $19.545  \pm0.170   $ & 16.283  \\
 0.0614  & $16.534  \pm0.102   $ & 21.221   & 1.5692  & $19.839  \pm0.141   $ & 16.212  \\
 0.0676  & $16.553  \pm0.114   $ & 21.221   & 1.7261  & $19.890  \pm0.243   $ & 16.149  \\
 0.0743  & $16.574  \pm0.126   $ & 20.628   & 1.8987  & $20.258  \pm0.141   $ & 16.087  \\
 0.0818  & $16.598  \pm0.135   $ & 20.628   & 2.0886  & $20.308  \pm0.232   $ & 16.024  \\
 0.0899  & $16.623  \pm0.146   $ & 20.628   & 2.2975  & $20.770  \pm0.147   $ & 15.968  \\
 0.0989  & $16.650  \pm0.159   $ & 20.628   & 2.5272  & $20.905  \pm0.220   $ & 15.917  \\
 0.1088  & $16.680  \pm0.156   $ & 20.281   & 2.7800  & $21.196  \pm0.162   $ & 15.866  \\
 0.1197  & $16.708  \pm0.148   $ & 19.809   & 3.0580  & $21.565  \pm0.252   $ & 15.819  \\
 0.1317  & $16.713  \pm0.143   $ & 19.809   & 3.3638  & $21.967  \pm0.094   $ & 15.785  \\
 0.1448  & $16.725  \pm0.132   $ & 19.638   & 3.7001  & $22.090  \pm0.274   $ & 15.751  \\
 0.1593  & $16.764  \pm0.117   $ & 19.254   & 4.0701  & $22.568  \pm0.117   $ & 15.720  \\
 0.1752  & $16.788  \pm0.102   $ & 19.254   & 4.4772  & $22.904  \pm0.102   $ & 15.695  \\
 0.1928  & $16.810  \pm0.090   $ & 19.060   & 4.9249  & $23.113  \pm0.115   $ & 15.673  \\
 0.2120  & $16.858  \pm0.075   $ & 18.837   & 5.4174  & $23.359  \pm0.147   $ & 15.649  \\
 0.2333  & $16.900  \pm0.054   $ & 18.652   & 5.9591  & $23.743  \pm0.096   $ & 15.622  \\
 0.2566  & $16.959  \pm0.049   $ & 18.418   & 6.5550  & $24.078  \pm0.101   $ & 15.604  \\
 0.2822  & $17.028  \pm0.057   $ & 18.341   & 7.2105  & $24.331  \pm0.099   $ & 15.580  \\
 0.3105  & $17.109  \pm0.074   $ & 18.150   & 7.9316  & $24.692  \pm0.082   $ & 15.560  \\
 0.3415  & $17.199  \pm0.076   $ & 17.991   & 8.7247  & $25.101  \pm0.116   $ & 15.545  \\
 0.3757  & $17.285  \pm0.066   $ & 17.835   & 9.5972  & $25.328  \pm0.118   $ & 15.534  \\
 0.4132  & $17.366  \pm0.060   $ & 17.662   & 10.5569 & $25.526  \pm0.127   $ & 15.518  \\
 0.4545  & $17.439  \pm0.064   $ & 17.532   & 11.6126 & $25.840  \pm0.164   $ & 15.496  \\
 0.5000  & $17.490  \pm0.074   $ & 17.382   & 12.7738 & $26.212  \pm0.170   $ & 15.483  \\
 0.5500  & $17.630  \pm0.072   $ & 17.249   & 14.0512 & $26.527  \pm0.176   $ & 15.469  \\
 0.6050  & $17.881  \pm0.058   $ & 17.108   & 15.4564 & $26.951  \pm0.233   $ & 15.462  \\
 \tableline
\end{tabular}
\end{center}
\end{table}
\begin{table}
\begin{center}
\caption{B514: Surface brightness, $\mu$, and integrated magnitude, $m$, as a function of the radius in the F814W filter of {\sl HST} ACS-WFC}
\begin{tabular}{ccc|ccc}
\tableline\tableline
   $R$    &  $\mu$                  &   $m$   &   $R$    &  $\mu$                 &   $m$  \\
 (arcsec) &  ($\rm{mag/arcsec^2}$)  &  (mag)  & (arcsec) &  ($\rm{mag/arcsec^2}$) &  (mag) \\
\tableline
 0.0260  & $15.810  \pm0.050   $ & 22.224   & 0.6655  & $17.332  \pm0.068   $ & 16.278  \\
 0.0287  & $15.819  \pm0.055   $ & 22.224   & 0.7321  & $17.451  \pm0.077   $ & 16.166  \\
 0.0315  & $15.828  \pm0.061   $ & 22.224   & 0.8053  & $17.601  \pm0.105   $ & 16.068  \\
 0.0347  & $15.839  \pm0.068   $ & 22.224   & 0.8858  & $17.837  \pm0.096   $ & 15.966  \\
 0.0381  & $15.850  \pm0.076   $ & 22.224   & 0.9744  & $18.048  \pm0.098   $ & 15.875  \\
 0.0420  & $15.862  \pm0.085   $ & 22.224   & 1.0718  & $18.007  \pm0.179   $ & 15.791  \\
 0.0461  & $15.874  \pm0.095   $ & 22.224   & 1.1790  & $18.194  \pm0.245   $ & 15.686  \\
 0.0508  & $15.888  \pm0.107   $ & 20.601   & 1.2969  & $18.447  \pm0.336   $ & 15.613  \\
 0.0558  & $15.903  \pm0.118   $ & 20.601   & 1.4266  & $18.755  \pm0.190   $ & 15.533  \\
 0.0614  & $15.917  \pm0.126   $ & 20.601   & 1.5692  & $19.082  \pm0.151   $ & 15.460  \\
 0.0676  & $15.931  \pm0.135   $ & 20.601   & 1.7261  & $19.078  \pm0.335   $ & 15.394  \\
 0.0743  & $15.947  \pm0.144   $ & 20.013   & 1.8987  & $19.444  \pm0.186   $ & 15.328  \\
 0.0818  & $15.965  \pm0.152   $ & 20.013   & 2.0886  & $19.540  \pm0.232   $ & 15.261  \\
 0.0899  & $15.982  \pm0.160   $ & 20.013   & 2.2975  & $19.965  \pm0.199   $ & 15.203  \\
 0.0989  & $15.999  \pm0.169   $ & 20.013   & 2.5272  & $20.101  \pm0.259   $ & 15.150  \\
 0.1088  & $16.015  \pm0.169   $ & 19.656   & 2.7800  & $20.402  \pm0.164   $ & 15.096  \\
 0.1197  & $16.034  \pm0.159   $ & 19.170   & 3.0580  & $20.718  \pm0.310   $ & 15.045  \\
 0.1317  & $16.024  \pm0.157   $ & 19.170   & 3.3638  & $21.153  \pm0.126   $ & 15.012  \\
 0.1448  & $16.022  \pm0.151   $ & 18.989   & 3.7001  & $21.253  \pm0.322   $ & 14.975  \\
 0.1593  & $16.055  \pm0.137   $ & 18.589   & 4.0701  & $21.778  \pm0.116   $ & 14.944  \\
 0.1752  & $16.072  \pm0.128   $ & 18.589   & 4.4772  & $22.117  \pm0.088   $ & 14.920  \\
 0.1928  & $16.078  \pm0.118   $ & 18.384   & 4.9249  & $22.297  \pm0.119   $ & 14.898  \\
 0.2120  & $16.120  \pm0.098   $ & 18.153   & 5.4174  & $22.484  \pm0.174   $ & 14.874  \\
 0.2333  & $16.171  \pm0.070   $ & 17.959   & 5.9591  & $22.916  \pm0.113   $ & 14.845  \\
 0.2566  & $16.225  \pm0.059   $ & 17.715   & 6.5550  & $23.213  \pm0.123   $ & 14.826  \\
 0.2822  & $16.287  \pm0.070   $ & 17.639   & 7.2105  & $23.470  \pm0.125   $ & 14.800  \\
 0.3105  & $16.367  \pm0.092   $ & 17.441   & 7.9316  & $23.880  \pm0.088   $ & 14.778  \\
 0.3415  & $16.454  \pm0.098   $ & 17.275   & 8.7247  & $24.263  \pm0.121   $ & 14.763  \\
 0.3757  & $16.558  \pm0.089   $ & 17.116   & 9.5972  & $24.457  \pm0.132   $ & 14.752  \\
 0.4132  & $16.646  \pm0.067   $ & 16.944   & 10.5569 & $24.607  \pm0.158   $ & 14.735  \\
 0.4545  & $16.708  \pm0.070   $ & 16.814   & 11.6126 & $24.935  \pm0.155   $ & 14.710  \\
 0.5000  & $16.740  \pm0.082   $ & 16.659   & 12.7738 & $25.294  \pm0.197   $ & 14.697  \\
 0.5500  & $16.878  \pm0.098   $ & 16.524   & 14.0512 & $25.649  \pm0.231   $ & 14.681  \\
 0.6050  & $17.137  \pm0.079   $ & 16.380   & 15.4564 & $26.034  \pm0.379   $ & 14.674  \\
\tableline
\end{tabular}
\end{center}
\end{table}
\begin{table}
\begin{center}
\caption{B514: Surface brightness profiles $\mu$ from star counts}
\begin{tabular}{cc|cc}
\tableline\tableline
   $R$    &  $\mu_{\rm{F606W}}$     &   $R$    &  $\mu_{\rm{F814W}}$          \\
 (arcsec) &  ($\rm{mag/arcsec^2}$)  & (arcsec) &        ($\rm{mag/arcsec^2}$) \\
\tableline
         &                           &   7.9625 & $   23.831 \pm   0.066 $ \\
  9.1875 & $   24.961 \pm    0.072 $ &   9.1875 & $   24.181 \pm   0.072 $ \\
 10.4125 & $   25.297 \pm    0.079 $ &  10.4125 & $   24.517 \pm   0.079 $ \\
 11.6375 & $   25.622 \pm    0.087 $ &  11.6375 & $   24.842 \pm   0.087 $ \\
 12.8625 & $   25.905 \pm    0.095 $ &  12.8625 & $   25.125 \pm   0.095 $ \\
 14.0875 & $   26.361 \pm    0.111 $ &  14.0875 & $   25.581 \pm   0.111 $ \\
 15.3125 & $   26.694 \pm    0.125 $ &  15.3125 & $   25.914 \pm   0.125 $ \\
 16.5375 & $   26.998 \pm    0.138 $ &  16.5375 & $   26.218 \pm   0.138 $ \\
 17.7625 & $   27.310 \pm    0.154 $ &  17.7625 & $   26.530 \pm   0.154 $ \\
 18.9875 & $   27.360 \pm    0.152 $ &  18.9875 & $   26.580 \pm   0.152 $ \\
 20.2125 & $   27.517 \pm    0.158 $ &  20.2125 & $   26.737 \pm   0.158 $ \\
 21.4375 & $   27.703 \pm    0.168 $ &  21.4375 & $   26.923 \pm   0.168 $ \\
 22.6625 & $   27.816 \pm    0.172 $ &  22.6625 & $   27.036 \pm   0.172 $ \\
 23.8875 & $   28.050 \pm    0.186 $ &  23.8875 & $   27.270 \pm   0.186 $ \\
 25.1125 & $   28.240 \pm    0.198 $ &  25.1125 & $   27.460 \pm   0.198 $ \\
 26.3375 & $   28.292 \pm    0.198 $ &  26.3375 & $   27.512 \pm   0.198 $ \\
 27.5625 & $   28.630 \pm    0.226 $ &  27.5625 & $   27.850 \pm   0.226 $ \\
 28.7875 & $   28.725 \pm    0.231 $ &  28.7875 & $   27.945 \pm   0.231 $ \\
 30.0125 & $   28.722 \pm    0.226 $ &  30.0125 & $   27.942 \pm   0.226 $ \\
 31.2375 & $   28.864 \pm    0.237 $ &  31.2375 & $   28.084 \pm   0.237 $ \\
 32.4625 & $   29.201 \pm    0.271 $ &  32.4625 & $   28.421 \pm   0.271 $ \\
 33.6875 & $   29.242 \pm    0.271 $ &  33.6875 & $   28.462 \pm   0.271 $ \\
 34.9125 & $   29.350 \pm    0.280 $ &  34.9125 & $   28.570 \pm   0.280 $ \\
 36.1375 & $   29.543 \pm    0.301 $ &  36.1375 & $   28.763 \pm   0.301 $ \\
 37.3625 & $   29.761 \pm    0.327 $ &  37.3625 & $   28.981 \pm   0.327 $ \\
 38.5875 & $   30.014 \pm    0.362 $ &  38.5875 & $   29.234 \pm   0.362 $ \\
 39.8125 & $   30.321 \pm    0.410 $ &  39.8125 & $   29.541 \pm   0.410 $ \\
 41.0375 & $   30.208 \pm    0.384 $ &  41.0375 & $   29.428 \pm   0.384 $ \\
\tableline
\end{tabular}
\end{center}
\end{table}
\begin{table}
\begin{center}
\caption{\label{Tab:assocation} Structural parameters of B514}
\begin{tabular}{lcc}
\tableline\tableline
Parameters         &             F606W                            &       F814W   \\
\tableline
$r_0$       & $0.36^{+0.09}_{-0.05}~\rm{arcsec}~(=1.35^{+0.35}_{-0.19}~\rm{pc})$      &
$0.36^{+0.09}_{-0.06}~\rm{arcsec}~(=1.35^{+0.32}_{-0.22}~\rm{pc})$ \\
$r_t$              & $16.08^{+2.11}_{-1.35}~\rm{arcsec}~(=61.11^{+8.00}_{-5.14}~\rm{pc})$    &
$16.79^{+1.74}_{-1.48}~\rm{arcsec}~(=63.78^{+6.62}_{-5.64}~\rm{pc})$    \\
$c=\log (r_t/r_0)$ &        $1.66^{+0.05}_{-0.04}$                 &   $1.68^{+0.04}_{-0.04}$ \\
$r_h$              &  $1.31^{+0.14}_{-0.08}~\rm{arcsec}~(=5.00^{+0.55}_{-0.32}~\rm{pc})$     &
$1.36^{+0.13}_{-0.09}~\rm{arcsec}~(=5.17^{+0.48}_{-0.34}~\rm{pc})$    \\
$\mu_0$ (${\rm mag~arcsec^{-2}}$)    &  $16.25^{+0.57}_{-0.56}$    &   $15.64^{+0.80}_{-0.64}$ \\
\tableline
\end{tabular}
\end{center}
\end{table}
\begin{table}
\begin{center}
\caption{BATC photometry of B514}
\begin{tabular}{ccccccc}
\tableline\tableline
Filter & Central wavelength & Bandwidth & Number of images & Exposure time &  rms      &  Magnitude \\
       &  (\AA)             &  (\AA)    &                  & (hours)       &  (mag)    &            \\
\tableline
$a$& 3360 & 222 & 6&  2:00 & 0.010 & $17.59\pm 0.05$\\
$b$& 3890 & 187 & 6&  2:00 & 0.010 & $16.82\pm 0.02$\\
$c$& 4210 & 185 & 4&  1:00 & 0.002 & $16.52\pm 0.01$\\
$d$& 4550 & 222 & 4&  1:20 & 0.015 & $16.25\pm 0.02$\\
$e$& 4920 & 225 & 3&  1:00 & 0.007 & $16.05\pm 0.01$\\
$f$& 5270 & 211 & 3&  1:00 & 0.014 & $15.85\pm 0.02$\\
$g$& 5795 & 176 & 3&  1:00 & 0.010 & $15.64\pm 0.01$\\
$h$& 6075 & 190 & 3&  0:50 & 0.005 & $15.56\pm 0.01$\\
$i$& 6660 & 312 & 3&  0:50 & 0.004 & $15.44\pm 0.01$\\
$j$& 7050 & 121 & 3&  1:00 & 0.006 & $15.33\pm 0.01$\\
$k$& 7490 & 125 & 3&  1:00 & 0.011 & $15.25\pm 0.01$\\
$m$& 8020 & 179 & 3&  1:00 & 0.003 & $15.19\pm 0.01$\\
$n$& 8480 & 152 & 3&  1:00 & 0.005 & $15.12\pm 0.01$\\
\tableline
\end{tabular}\\
\end{center}
\end{table}
\begin{table}
\begin{center}
\caption{2MASS and $GALEX$ photometry of B514} \label{t8.tab}
\begin{tabular}{cc}
\tableline\tableline
Filter & Magnitude \\
\tableline
$J$    & $14.23\pm 0.07$\\
$H$    & $13.32\pm 0.06$\\
$K_s$  & $13.63\pm 0.10$\\
NUV    & $19.57\pm 0.86$\\
FUV    & $20.44\pm 2.30$\\
\tableline
\end{tabular}\\
\end{center}
\end{table}
\begin{table}
\begin{center}
\caption{Mass estimates (and uncertainties) of B514 based on the BC03 models} \label{t9.tab}
\begin{tabular}{ccccccccc}
\tableline\tableline
 $V$ &  $u$ &  $g$ &  $r$ &  $i$ &  $z$ &  $J$ & $H$ & $K_{\rm s}$      \\
\tableline
\multicolumn{1}{c}{} & \multicolumn{7}{c}{Mass $(10^6~M_\odot)$} & \multicolumn{1}{c}{}\\
\tableline
$1.08\pm0.03$&$0.99\pm0.03$ & $1.06\pm0.03$ & $1.02\pm0.03$ & $1.0\pm0.03$ & $0.98\pm0.03$ & $0.98\pm0.06$ & $1.41\pm0.08$ &$0.96\pm0.09$  \\
\tableline
\end{tabular}
\end{center}
\end{table}

\end{document}